\RequirePackage{fix-cm}
\documentclass[smallextended]{svjour3}       % onecolumn (second format)
\smartqed  % flush right qed marks, e.g. at end of proof
\usepackage{graphicx}
\usepackage[misc]{ifsym}
\usepackage{authblk}
\usepackage{subcaption}
\begin{document}

\title{Improving the Energy Resolution of the Reactor Antineutrino Energy Reconstruction with Positron Direction%\thanks{Grants or other notes
}

\author{Lianghong Wei\textsuperscript{1,2}
\and Liang Zhan\textsuperscript{1}
\and Jun Cao\textsuperscript{1,2}
\and Wei Wang\textsuperscript{3}
}

\institute{
\Letter Liang Zhan\\
\email{zhanl@ihep.ac.cn}\\
\at
 $^{1}$~Institute of High Energy Physics, Chinese Academy of Sciences, Beijing 100049, China
\at
 $^{2}$~University of Chinese Academy of Sciences, Beijing 100049, China
\at
 $^{3}$~Nanjing University, Nanjing, 210093, China\\
}

\titlerunning{Improving the Energy Resolution of $\bar\nu_e$ with Positron ...}  % if too long for running head

\authorrunning{Lianghong Wei, Liang Zhan, Jun Cao, Wei Wang} % if too long for running head

\date{Received: date / Accepted: date}

\maketitle
\begin{abstract}
The energy resolution is crucial for the reactor neutrino experiments which aims to determine neutrino mass ordering by precise measurement of the reactor antineutrino energy spectrum.
A non-negligible effect in the antineutrino energy resolution is the spread of the kinetic energy of the recoiled neutron and the corresponding positron when detecting the antineutrinos via Inverse Beta-Decay (IBD) reaction. The emission direction of the produced positron in IBD reaction can be used to estimate the kinetic energy of neutron and thus the reconstructed antineutrino energy resolution can be improved.
To demonstrate the feasibility, a simple positron direction reconstruction method is implemented in a toy liquid scintillator detector like the Taishan Antineutrino Observatory (TAO) with 4500 photoelectron yield per MeV.
A 4\% to 26\% improvement of energy resolution can be achieved for 5 MeV reactor antineutrinos at TAO.

\keywords{energy resolution \and neutron recoiling \and positron direction reconstruction  \and Cerenkov}
\end{abstract}

\section{Introduction}

The Neutrino oscillation phenomena opens a door to new physics beyond the Standard Model of particle physics. Since 1998, a number of atmospheric, solar, accelerator and reactor experiments have provided us with very compelling evidences for neutrino oscillations. The ongoing and future neutrino oscillation experiments are expected to probe the neutrino mass ordering and the value of CP violating phase. Jiangmen Underground Neutrino Observatory (JUNO)~\cite{Djurcic:2015vqa,An:2015jdp} is proposed to determine the neutrino mass ordering with precise measurement of the reactor antineutrino energy spectrum. The energy resolution is crucial in order to determine neutrino mass ordering. The JUNO detector has 20-kton liquid scintillator as detection target with a designed energy resolution of 3\%/$\sqrt{E( MeV)}$.
Recent reactor neutrino experiments, Daya Bay~\cite{An:2015nua,An:2016srz,Adey:2019ywk}, Double Chooz~\cite{Abe:2014bwa}, RENO~\cite{Seon-HeeSeofortheRENO:2014jza}, and  NEOS~\cite{Ko:2016owz} have shown that the theoretical reactor antineutrino energy spectrum disagree with the observed energy spectrum.
Furthermore, the antineutrino energy spectrum shows fine structures in the summation of the spectra of thousands of beta-decay branches of fission products~\cite{Dwyer:2014eka}.
To provide a high precision reference spectrum for JUNO, Taishan Antineutrino Observatory (TAO)~\cite{TAO_CDR} is proposed as a satellite experiment of JUNO with an energy resolution better than  $2\%/\sqrt{E( MeV)}$. TAO will also provide a high precision and high energy resolution measurement of the reactor antineutrino spectrum as a benchmark to test nuclear databases~\cite{INDC-NDS-0786}.

The reactor antineutrinos are usually detected by IBD reaction,
$\bar\nu_e \ + \ p \rightarrow e^+ \ + \ n $, in liquid scintillator detectors.
The positron kinetic energy is a good approximation of the incident antineutrino energy with an approximate shift of 1.8~MeV.
The kinetic energy of the neutron spreads in a range of zero to tens of keV, and has an impact on the energy resolution at sub-percent level in the determination of the antineutrino energy.
This effect is negligible at experiments with large energy resolution, such as $8\%/\sqrt{E( MeV)}$ at Daya Bay experiment.
However it becomes non-trivial when improving the energy resolution to the level of $2\%/\sqrt{E( MeV)}$, as the design goal of the TAO experiment.
In this paper, we propose a method to improve the energy resolution by reducing the impact from the spread of the neutron kinetic energy.
The neutron kinetic energy can be determined to correct the antineutrino energy with the information of the positron scattering direction.
In a real detector, the reconstruction of the positron scattering direction provides information of the neutron kinetic energy with a certain resolution and improves the energy resolution.

\section{Energy resolution due to neutron recoiling}

The energy resolution of a reactor antineutrino detector using liquid scintillator is dominated by the statistical fluctuation of the number of the collected photoelectrons on photosensors.
In JUNO and TAO experiments, the photoelectron yield per MeV energy is about 1200 and 4500~\cite{TAO_CDR}, respectively.
The corresponding energy resolution is about 3\% and 1.5\% at 1~MeV by photoelectron statistics.
As a secondary effect, the neutron recoiling can affect the energy resolution at sub-percent level.

%The kinetic energy of neutron in IBD reaction shares a part of initial energy of the incident antineutrino, which introduces energy spread of positron and neutron in the final state.
In IBD reaction, the reactor antineutrino~($\bar{\nu}_e$) interacts with a proton, creating a positron~(e${^+}$) and a neutron.
The energy relation is
\begin{equation}
E_{\bar{\nu}_e} = T_{e^{+}} + T_n + \Delta_{np} + m_{e^+} \,, \label{eq:IBD}
\end{equation}
where $E_{\bar{\nu}_e}$ is the incident antineutrino energy, $T_{e^{+}}$ is the positron kinetic energy, $T_n$ is the neutron kinetic energy, $\Delta_{np}$ is the mass difference of neutron and proton, and $m_{e^+}$ is the positron mass.
$T_n$ spreads from 0 to a few tens of keV. When neglecting $T_n$, the positron kinetic energy is a approximation of antineutrino energy by a shift of 1.8~MeV, $E_{\bar{\nu}_e} \approx T_{e^{+}} + 1.8~\mathrm{MeV}$.

\begin{figure}[htb]
\centering
\includegraphics[scale=0.5]{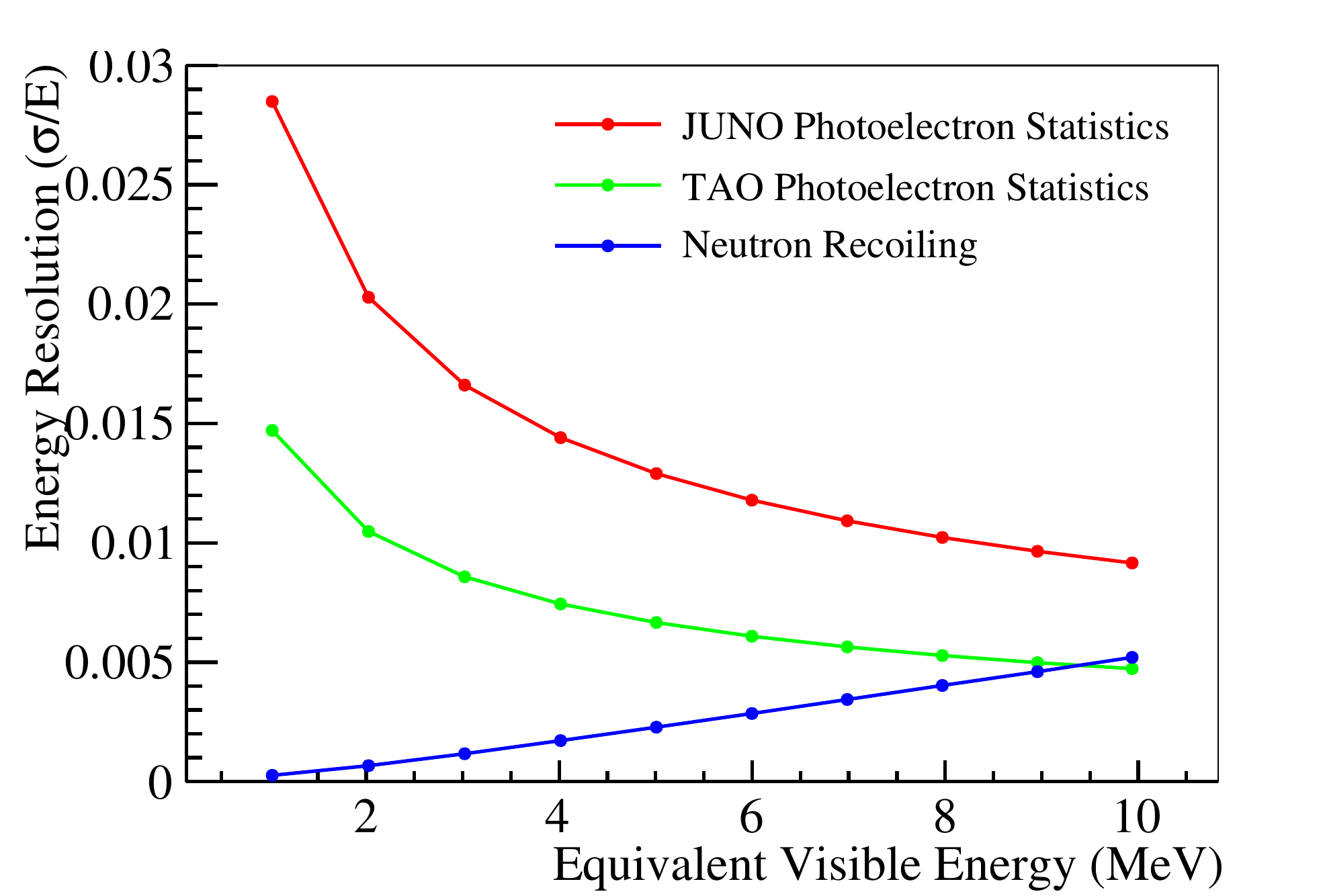}
\caption{Energy resolution for the equivalent visible energy defined as the antineutrino energy minus a constant energy shift of 0.78~MeV due to the IBD reaction kinetics. Red (green) line is the energy resolution from photoelectron statistics for the JUNO (TAO) experiment. Blue line is the energy resolution from neutron recoiling with liquid scintillator quenching taken into account.}
\label{fig:energy resolution from statistic and neutron recoiling}
\end{figure}

For IBD reaction, the positron angular distribution was calculated by Petr Vogel~\cite{Vogel:1999zy}.
Based on the distribution, the energy spread of the positron and neutron is $\Delta E =2(E_{\bar{\nu}_e}-\Delta_{np})E_{\bar{\nu}_e}/M_p$, with $M{_p}$ the proton mass.
The calculated neutron kinetic energy with a given antineutrino energy follows approximately uniform distribution with a spread of $\Delta E$.
The induced energy smearing for recoiled neutron is approximately $\Delta E/\sqrt{12}$, as the standard deviation of the uniform distribution.
As shown in Fig.~\ref{fig:energy resolution from statistic and neutron recoiling}, 
the energy resolution for the equivalent visible energy is calculated. The equivalent visible energy is defined as the antineutrino energy minus a constant energy shift of 0.78~MeV due to the IBD reaction kinetics.
The energy resolution from neutron recoiling is much smaller than the photoelectron statistics for JUNO as the total energy resolution is the square root of the quadratic sum of the two effects.
However, the energy resolution from neutron recoiling can even be larger than that from the photoelectron statistics around 10~MeV for TAO.

The reactor antineutrino detector detects the positron energy to obtain the information of the antineutrino energy.
The positron predominantly deposits its kinetic energy and annihilates into two 0.511~MeV gammas, which gives a prompt signal.
The neutron scatters in the detector until being thermalized and then it is captured to produce a delayed signal.
The visible energy ($E_{vis}$) of prompt signal of IBD in liquid scintillator can be calculated via
\begin{equation}
E_{vis} = T_{e^{+}} + 2 \times 0.511 + Q_F \times T_n \,, \label{con:eq1}
\end{equation}
where $Q_F$ is the neutron quenching factor defined as the light yield ratio of neutron to electron in liquid scintillator.
When neutron recoils on the protons, the energetic protons can generate a small amount of light which could be mixed with the light generated by the positron.
The few-keV neutron kinetic energy could contribute a small correction, $Q_F \times T_n$, to the prompt energy.

\begin{figure}[htb]
\centering
\includegraphics[scale=0.5]{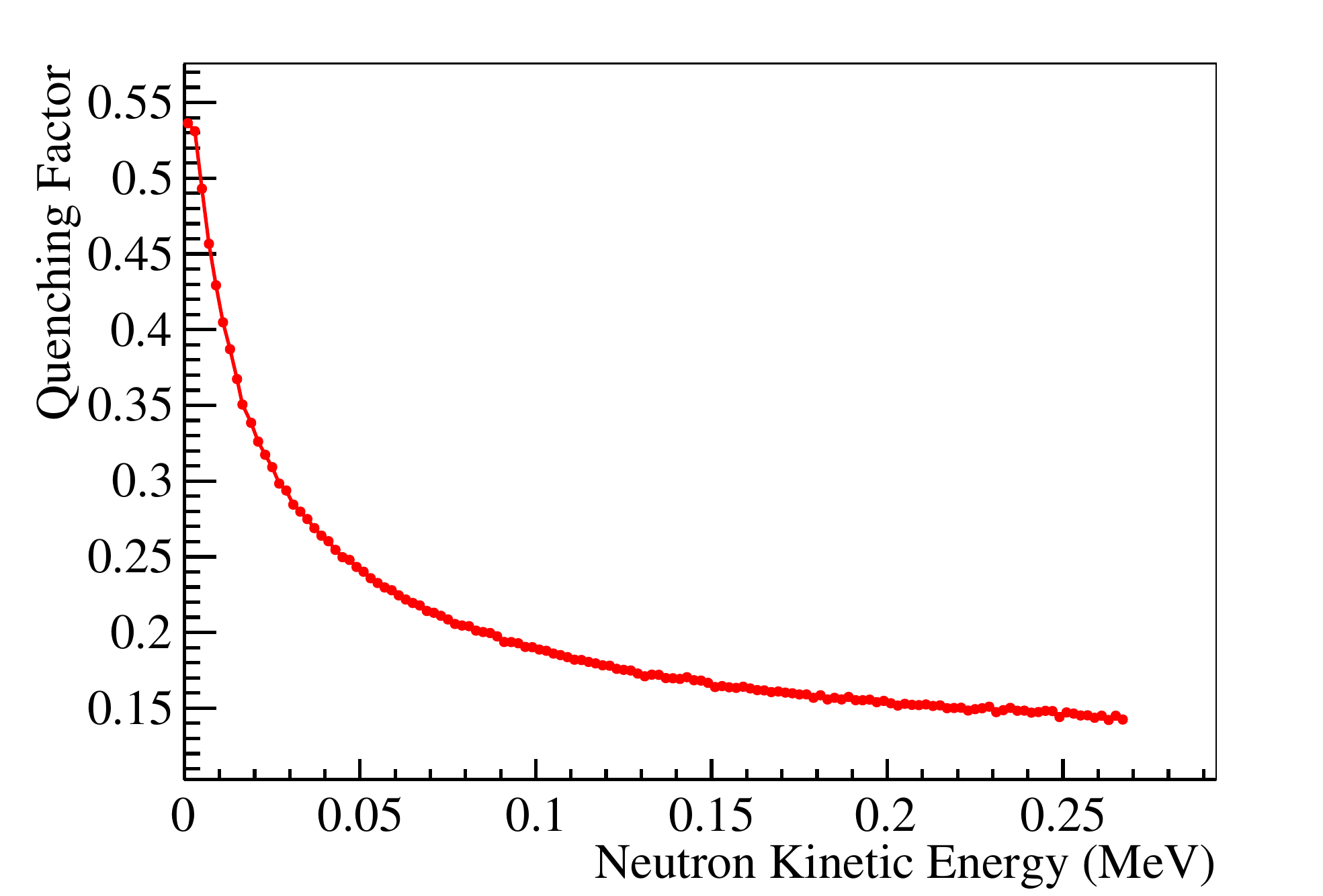}
\caption{The neutron quenching factor at different energies is obtained by Geant4 simulation.}
\label{fig:QF_curve}
\end{figure}

One has to take into account that the light output from the recoil proton is quenched.
The quenching mechanism was first discussed by Birks~\cite{Birks:1964zz}.
The light output of the liquid scintillator is related to the energy deposition density $dE/dr$.
An empirical model is commonly used to describe the process,
\begin{equation}
 \frac{dL}{dr} = S\frac{\frac{dE}{dr}}{1 + k_B\frac{dE}{dr} + k_C(\frac{dE}{dr})^{2}} \,, \label{con:quenching}
\end{equation}
where $dL/dr$ is the scintillation light yield per unit path length $r$, $S$ is the the scintillation light yield per MeV, $dE/dr$ is the energy deposition density, $k_B$ is the Birks’ constant, and $k_C$ is the second order parameter. The Birks’ constants,
$k_B = 6.5\times 10^{-3}$~g/cm$^{2}$/MeV and $k_C = 1.5 \times 10^{-6}$~g$^2$/cm$^4$/MeV$^2$, for Linear Alkylbenzen (LAB) based liquid scintillator  are taken from Ref.~\cite{An:2015jdp}, and will be used for JUNO and TAO.
It should be noted that the Birks’ constants rely on the different types of liquid scintillator, and also rely on the modelling in Monte Carlo because $dE/dr$ is not directly visible. For the same quenching effect, different modelling in Monte Carlo may result in different Birks' constants. 
The values of the quenching factors of the gamma and positron have negligible impact on the energy resolution due to neutron recoiling and is approximately set to be 1.0 in the calculation.  
We simulate neutrons with different energies and obtain a curve of $Q_F$ as shown in Fig.~\ref{fig:QF_curve}.

Given the positron energy and the positron scattering angle, based on conservation of energy and momentum in IBD reaction, the kinetic energy of the neutron can be calculated as

\begin{equation}
E_{n} = \frac{E_{e^+} \times M_{p} +  (E_{e^+} - M_{p})  \times P_{e^+} \times  \cos{\theta} - {E_{e^+}}^2 - C}{E_{e^+} - M_{p} - P_{e^+} \times \cos{\theta}} \\, \label{con:neutron ke}
\end{equation}
where $C=0.5 \times ({M_{p}}^2+{M_{n}}^2-{m_{e^+}}^2)$, $E_{e{^+}}$ is the energy of positron which can be gotten by measuring $E_{vis}$ using relation in Eq.~\ref{con:eq1}, and $P_{e^+}$ is the momentum of the positron.
If the positron scattering angle ($\theta$ or $\cos\theta$) is obtained precisely, the kinetic energy of neutron can be determined by Eq.~\ref{con:neutron ke} and energy smearing due to neutron recoiling will be removed.
For reactor neutrino experiments, when both the detector size and the reactor core size can be ignored in comparison to the distance between them, the antineutrino direction is assumed to be known.
Positron direction reconstruction is the key to obtain the scattering angle to improve the energy resolution by reducing the neutron recoiling smearing.

\section{Positron direction reconstruction}

The basic idea of the positron direction reconstruction is to utilize the direction of Cerenkov light in liquid scintillator.
The axis of Cerenkov cone can be used to present the positron direction.
The refractive index of liquid scintillator of JUNO and TAO is 1.5 at 430~nm.
Thus the Cerenkov radiation threshold for positron is 0.174~MeV.
The scintillation light is isotropic, and no direction information can be derived.
Therefore, the scintillation light is a type of background when using the Cerenkov light to reconstruct the positron direction.
The key point for the direction reconstruction is to select as more as possible Cerenkov photons in the sea of scintillation photons.
One possible way to distinguish the Cerenkov light and the scintillation light is to use their hit time difference~\cite{Aberle:2013jba} on photosensors.
Scintillator light usually has fast and slow components, while the Cerenkov light is emitted immediately because of its luminescence mechanism. Cerenkov light with short wavelength will be absorbed by the liquid scintillator and remitted as scintillation light. A fraction of long wavelength Cerenkov light will survive and could dominate in the earliest hits on photosensors.

Borrowing the experience of JUNO simulation software~\cite{Djurcic:2015vqa}, a standalone Geant4 simulation package is developed for the TAO simulation~\cite{TAO_CDR}. A brief description of the parameters in simulation is described as follows. The liquid scintillator is contained in an acrylic vessel of a diameter of 1.8~m. About 4100 Silicon Photomultiplier (SiPM) tiles, with a dimension of $50\times50$~mm$^2$ and a photon detection efficiency (PDE) of 50\%, are placed just 2-cm away from the acrylic vessel as photosensors.
The liquid scintillator properties in the simulation are taken from Ref.~\cite{An:2015jdp}. The time constants of the liquid scintillator are essential for the hit time distribution and are listed in Table~\ref{tab:1}.
\begin{table}[htb]
\caption{The time constants and the corresponding fractions of the three components of the liquid scintillator for three species of particles.}
\label{tab:1}       % Give a unique label
\begin{tabular}{llll}
\hline\noalign{\smallskip}
Particles  &  Fast~(ns)/ratio  &  Slow~(ns)/ratio  &  Slower~(ns)/ratio \\
\noalign{\smallskip}\hline\noalign{\smallskip}
      $\gamma$,e$^-$,e$^+$   &  4.93/79.9\%  & 20.6/17.1\%  & 190/3.0\% \\
      n,p                    &  4.93/65\%    & 34.0/23.1\%  & 220/11.9\% \\
      $\alpha$               &  4.93/65\%    & 35.0/22.8\%  & 220/12.2\%\\
\noalign{\smallskip}\hline
\end{tabular}
\end{table}

\begin{figure}[htb]
\centering
\includegraphics[scale=0.5]{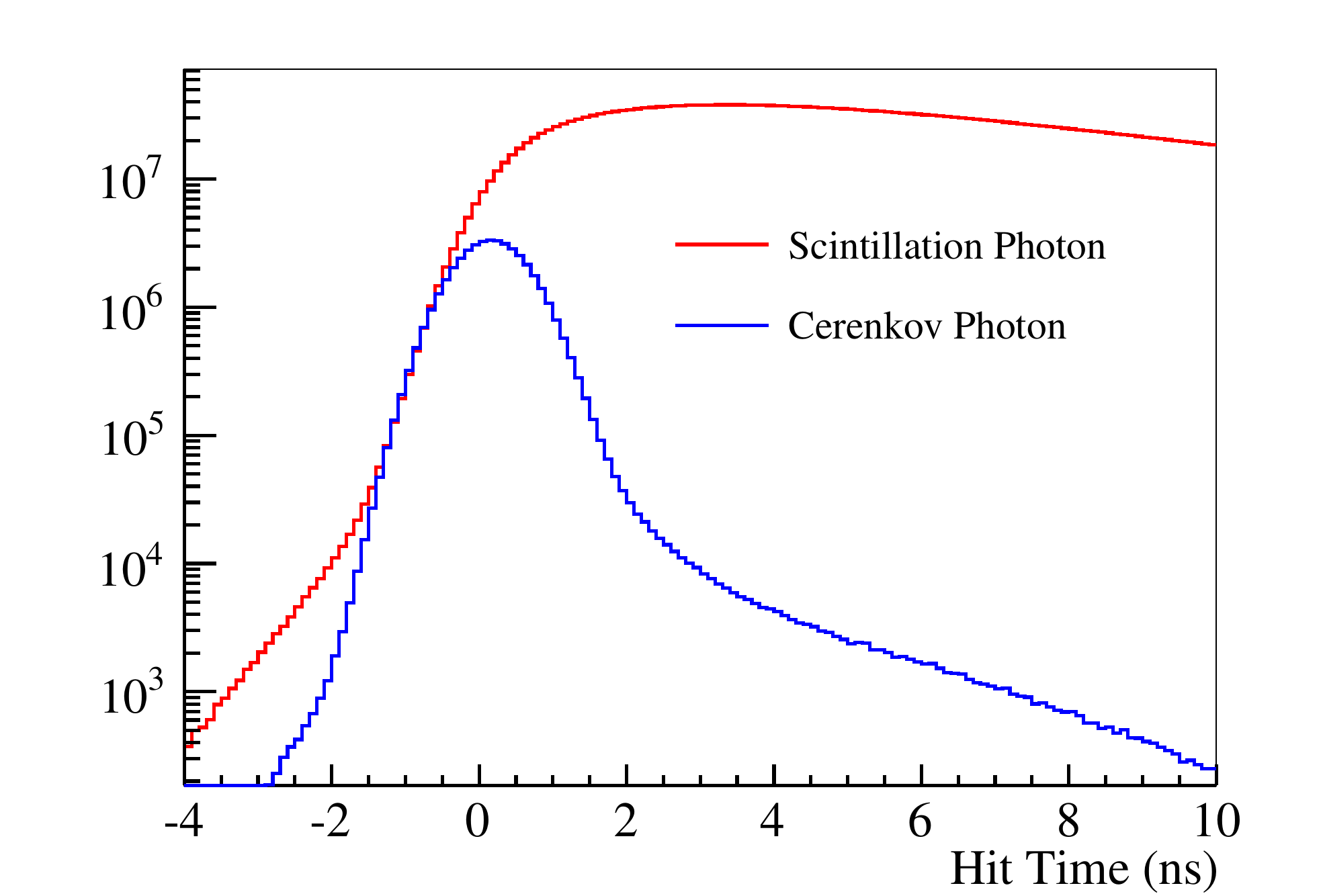}
\caption{The hit time distribution of the Cerenkov and scintillation photons with 0.5~ns time resolution after correction of the time of flight. The time zero is defined as the generation time of the IBD event.}
\label{fig:hit_time}
\end{figure}

The positron direction reconstruction are performed including two steps.
The first step is to obtain the hit time distribution after correcting the time of flight with the position of the reconstructed event vertex and the location of photosensors.
Accurate vertex reconstruction and a good timing resolution of electronics are crucial for selecting the Cerenkov photons.
A 2-cm vertex resolution for reactor antineutrinos can be obtained for TAO based on a simple vertex algorithm.
This algorithm uses the gravity center of charges of all SiPMs and a simulated correction map of the gravity center to the true vertex. 
For the time resolution of each readout channel, the time resolution of 0~ns, 0.5~ns, and 1~ns are assumed respectively and used to smear the hit time.
The typical hit time distribution is shown in Fig.~\ref{fig:hit_time} for both the scintillation and the Cerenkov photons.
The direction reconstruction performance depends on the total number of photons and the purity of the Cerenkov photons in the selection window.
After optimization, we use a time cut of "$<0$~ns" in the time distribution to select a sample with high purity of Cerenkov photons.

\begin{figure}[htb]
\centering
\includegraphics[scale=0.5]{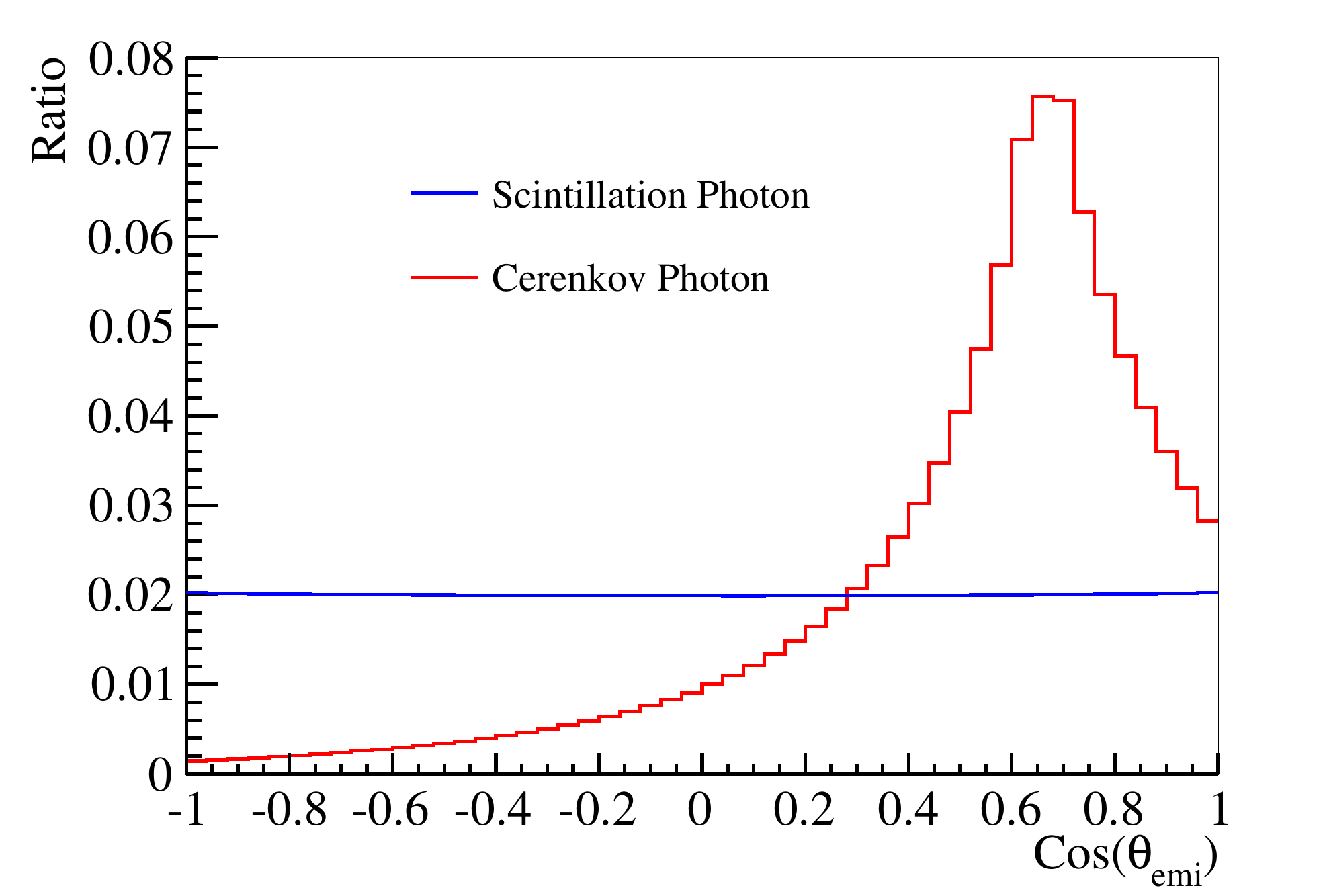}
\caption{Cerenkov photons and scintillation photons emission angles distribution after normalization.}
\label{fig:photon emission angle}
\end{figure}

The second step is the direction reconstruction using the selected photons. Since the Cerenkov radiation forms a cone centered on the particle moving direction, the direction can be determined by taking the centroid of all vectors pointing from the reconstructed event vertex to the position of the hit photosensors~\cite{Cheng:2015cen}. In Eq.~\ref{con:eq4}, $\vec{D}$ is the reconstructed direction of positron, $X_{positron}$ is the positron reconstructed vertex, $j$ is the identifier (ID) of the hit photosensors, and $q^{j}$ and ${\vec{X}_i}^j$ are the charge and position of the hit photosensor, respectively.
\begin{equation}
\vec{D} = \sum\limits_{j}^{N} q^j({\vec{X}}^j-\vec{X_{positron}})   \label{con:eq4}
\end{equation}
Because of the multi-scattering, the Cerenkov ring is fuzzy since positron changes its direction during the ionization process. The Cerenkov photon emission angle distribution is displayed in Fig.~\ref{fig:photon emission angle}.
The isotropic scintillation light is a severe background to the Cerenkov light.
To study the effects of the scintillation light pollution and the impact of the time resolution, we define an intrinsic angular resolution with the angle between the true and the reconstructed direction of positron without the pollution of the scintillation light and the impact of the hit time resolution. That is, only the statistics and the spread of Cerenkov light due to multi-scattering and the vertex smearing are considered. In Monte Carlo, only the Cerenkov hits are selected to reconstruct the direction.

\begin{figure}[htb]
\begin{subfigure}{.5\textwidth}
  \centering
  % include first image
  %\includegraphics[width=.8\linewidth]{direction_res_update2.pdf}  
  \includegraphics[scale=0.3]{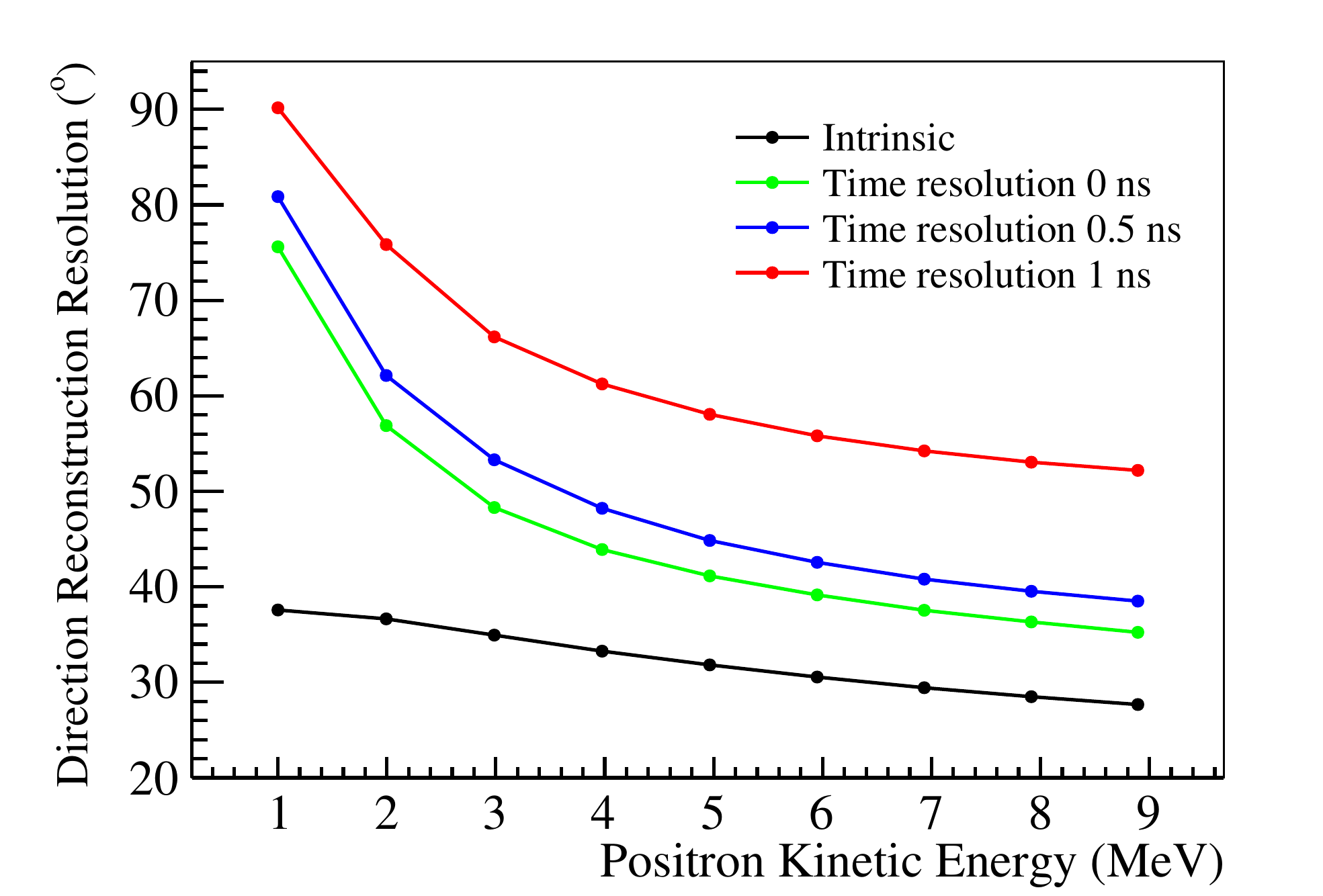}
  \caption{}
  \label{fig:sub-first}
\end{subfigure}
\begin{subfigure}{.5\textwidth}
  \centering
  % include second image
  %\includegraphics[width=.8\linewidth]{theta_res_update2.pdf}
  \includegraphics[scale=0.3]{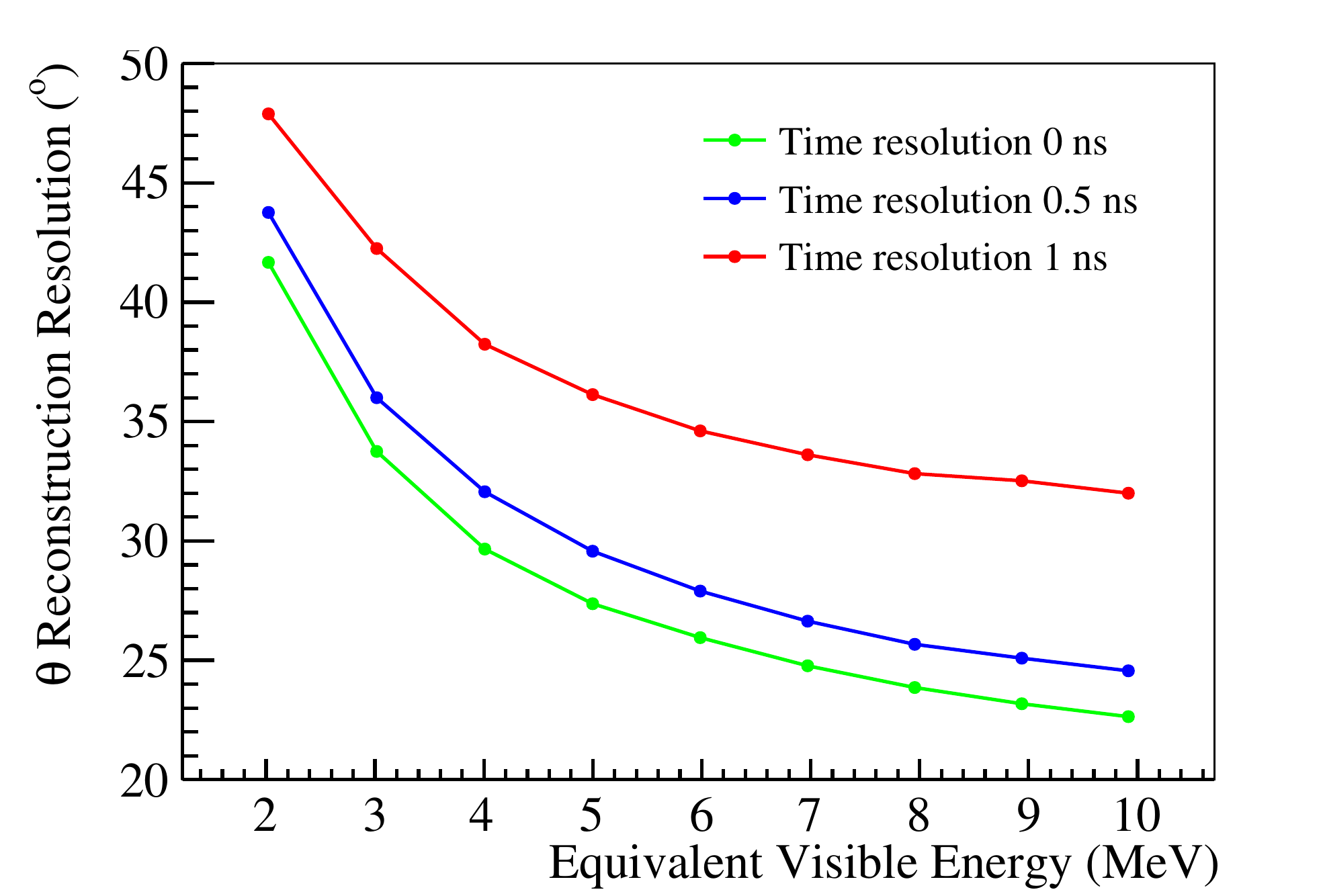} 
  \caption{}
  \label{fig:sub-second}
\end{subfigure}
\caption{The resolution of positron direction reconstruction is defined as 68.3\% of the reconstructed directions contained in a cone centered by the true direction within this angle.
(a) shows the positron direction resolution with different time resolution and black line is the intrinsic angular resolution of positron; (b) shows positron scattering angular ($\theta$) reconstruction resolution with different equivalent visible energy.}
\label{fig:fig}
\end{figure}

To characterize the performance of the positron direction reconstruction, the resolution of the reconstruction is defined as 68.3\% of the reconstructed directions contained in a cone centered by the true direction within this angle. The resolution relies on the timing resolution of electronics as shown in Fig.~\ref{fig:sub-first}. The intrinsic resolution has been defined above. Adding back the scintillation light pollution, the resolution worsen as shown by the green curve, labelled as "Time resolution 0~ns". Considering the time resolution of the hit time of 0.5 and 1~ns, the angular resolution further degrade as shown by the blue and red curve, respectively.

\section{Improved energy resolution from neutron recoiling}
\label{sec:5}
To determine the neutron kinetic energy using Eq.~\ref{con:neutron ke}, we calculate the positron scattering angle ($\theta$) based on the reconstructed positron direction. The scattering angle resolution is evaluated as the standard deviation of the difference of the reconstructed $\theta$ and true $\theta$. Fig.~\ref{fig:sub-second} shows the scattering angle resolution as a function of equivalent visible energy.
The scattering angle resolution is propagated to the neutron kinetic energy spread.

\begin{figure}[htb]
\centering
\includegraphics[scale=0.5]{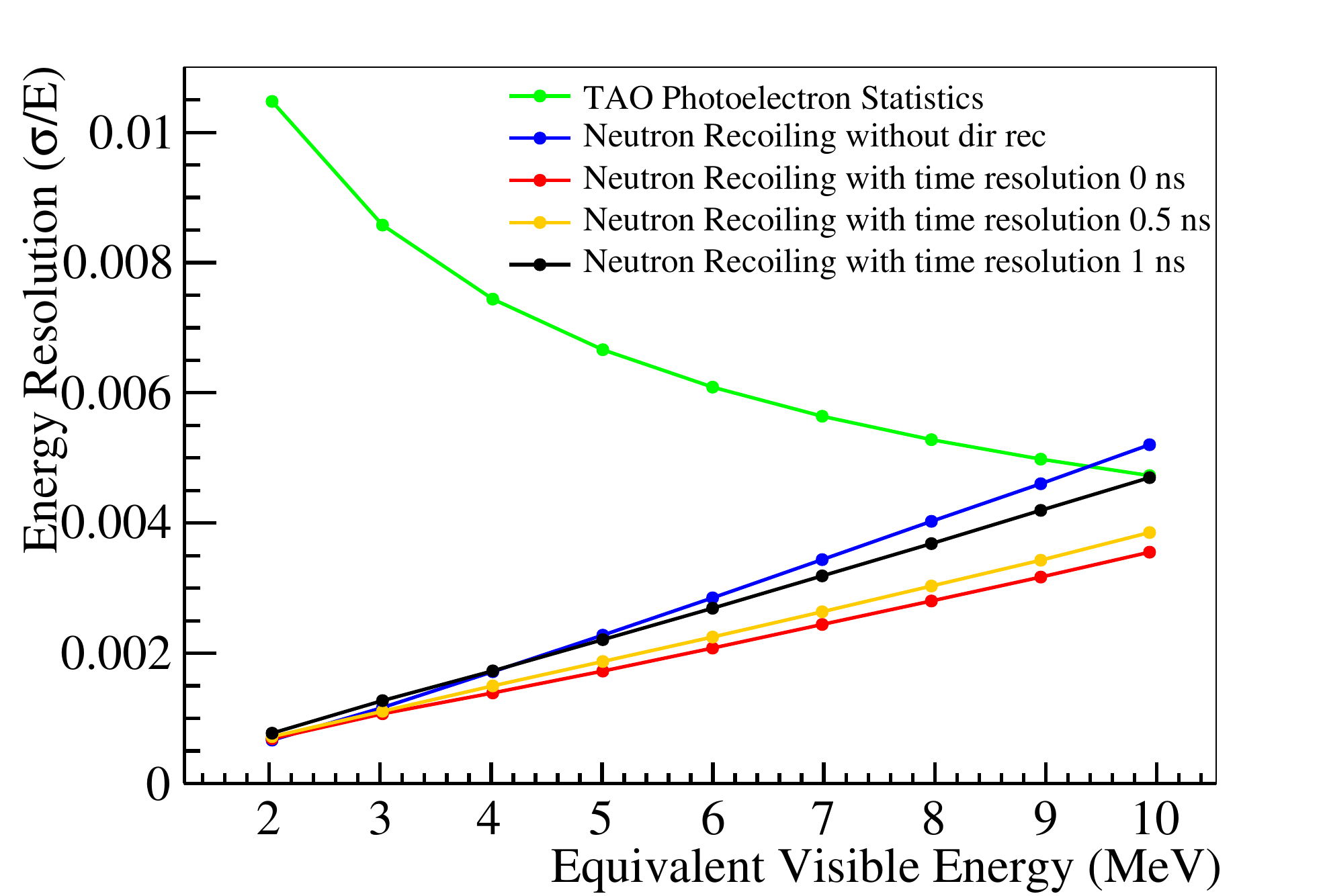}
\caption{Energy resolution for equivalent visible energy. Green line is the energy resolution from photoelectron statistics for the TAO experiment. Blue line is the energy resolution from neutron recoiling with a factor~(1-$Q_F$). The others are the energy resolution from neutron recoiling after positron direction reconstruction with different time resolution.}
\label{fig:energy resolution}
\end{figure}

Based on Eq.~\ref{eq:IBD} and Eq.~\ref{con:eq1}, the neutrino reconstruction energy ($E_{rec}$) can be calculated with the detected visible energy $E_{vis}$ and correction from neutron kinetic energy.
\begin{equation}
E_{rec} = E_{vis} - 0.511 + \Delta _{np} + (1-Q_F) \times T_{n} \label{con:eq5}
\end{equation}
The resolution of $E_{rec}$ due to neutron recoiling can be calculated with the resolution of $T_n$ and a factor of $(1-Q_F)$. With positron direction reconstruction, the resolution of $T_n$ relies on $\theta$ reconstruction resolution and is better than that of the original spread without information of positron direction.
For examples, for 5 MeV reactor antineutrinos, without positron direction reconstruction, the spread of neutron kinetic energy is $(1-Q_F) \times \Delta E/\sqrt{12}$ and its contribution to the energy resolution of IBD positron signal is 0.23\%. Considering positron direction reconstruction, the resolution becomes 0.17\%~(0.22\%) with 0~ns~(1~ns) time resolution, so the energy resolution of IBD positron signal can be improved by 4\% to 26\% which depends on the time resolution.
Fig.~\ref{fig:energy resolution} updates the energy resolution of antineutrino compared with Fig.~\ref{fig:energy resolution from statistic and neutron recoiling}.
With 0.5~ns time resolution, the energy resolution from neutron recoiling is obviously improved. For TAO, it is smaller than the resolution caused by the statistical fluctuation of collected photoelectrons.
For the worst case to vary the $Q_F$ value within 30\%, the energy resolution is still obviously improved with the help from positron direction reconstruction at a level of 0\%~-~22\% depending on various time resolutions from 1~ns to 0~ns. 

\section{Conclusion}
\label{sec:conc}
For reactor neutrino experiments detecting the antineutrinos via IBD reaction, the energy resolution is crucial in order to determine neutrino mass ordering with precise measurement of the reactor antineutrino energy spectrum. The spread of the kinetic energy of the recoiled neutron is a non-negligible effect in the energy resolution of antineutrino and can be significantly improved by the direction reconstruction of the produced positron in IBD reaction. A simple positron direction reconstruction method is implemented in a toy liquid scintillator detector with 4500 photoelectron yield per MeV like TAO. A 4\% to 26\% improvement of energy resolution could be achieved for 5~MeV reactor antineutrinos.

\begin{acknowledgements}
This work is supported by the National Natural Science Foundation of China under Grant No.11775247 and the National Key R\&D Program of China under Grant No.2018YFA0404100

\end{acknowledgements}

% could have direct or potential influence or impart bias on.


\bibliographystyle{unsrt}
\bibliography{references}

\begin{thebibliography}{10}

\bibitem{Djurcic:2015vqa}
Zelimir Djurcic et~al.
\newblock {JUNO Conceptual Design Report}.
\newblock 2015.

\bibitem{An:2015jdp}
Fengpeng An et~al.
\newblock {Neutrino Physics with JUNO}.
\newblock {\em J. Phys.}, G43(3):030401, 2016.

\bibitem{An:2015nua}
Fengpeng An et~al.
\newblock {Measurement of the Reactor Antineutrino Flux and Spectrum at Daya
  Bay}.
\newblock {\em Phys. Rev. Lett.}, 116(6):061801, 2016.
\newblock [Erratum: Phys. Rev. Lett. 118, no.9, 099902 (2017)].

\bibitem{An:2016srz}
Fengpeng An et~al.
\newblock {Improved Measurement of the Reactor Antineutrino Flux and Spectrum
  at Daya Bay}.
\newblock {\em Chin. Phys.}, C41(1):013002, 2017.

\bibitem{Adey:2019ywk}
D.~Adey et~al.
\newblock {Extraction of the $^{235}$U and $^{239}$Pu Antineutrino Spectra at
  Daya Bay}.
\newblock {\em Phys. Rev. Lett.}, 123(11):111801, 2019.

\bibitem{Abe:2014bwa}
Y.~Abe et~al.
\newblock {Improved measurements of the neutrino mixing angle $\theta_{13}$
  with the Double Chooz detector}.
\newblock {\em JHEP}, 10:086, 2014.
\newblock [Erratum: JHEP02, 074 (2015)].

\bibitem{Seon-HeeSeofortheRENO:2014jza}
Seon-Hee Seo.
\newblock {New Results from RENO and The 5 MeV Excess}.
\newblock {\em AIP Conf. Proc.}, 1666(1):080002, 2015.

\bibitem{Ko:2016owz}
Y.~J. Ko et~al.
\newblock {Sterile Neutrino Search at the NEOS Experiment}.
\newblock {\em Phys. Rev. Lett.}, 118(12):121802, 2017.

\bibitem{Dwyer:2014eka}
D.~A. Dwyer and T.~J. Langford.
\newblock {Spectral Structure of Electron Antineutrinos from Nuclear Reactors}.
\newblock {\em Phys. Rev. Lett.}, 114(1):012502, 2015.

\bibitem{TAO_CDR}
Angel Abusleme et~al.
\newblock {TAO Conceptual Design Report}.
\newblock 2020.

\bibitem{INDC-NDS-0786}
M.~Fallot, B.~Littlejohn, and P.~Dimitriou.
\newblock {Antineutrino spectra and their applications}, 2019.
\newblock {International Atomic Energy Agency Report INDC(NDS)-0786 (2019)}.

\bibitem{Vogel:1999zy}
P.~Vogel and John~F. Beacom.
\newblock {Angular distribution of neutron inverse beta decay, anti-neutrino(e)
  + p $\rightarrow e^+$ + n}.
\newblock {\em Phys.\ Rev.\ D}, 60:053003, 1999.

\bibitem{Birks:1964zz}
John~B. Birks.
\newblock {\em {The Theory and practice of scintillation counting}}.
\newblock 1964.

\bibitem{Aberle:2013jba}
C.~Aberle, A.~Elagin, H.~J. Frisch, M.~Wetstein, and L.~Winslow.
\newblock {Measuring Directionality in Double-Beta Decay and Neutrino
  Interactions with Kiloton-Scale Scintillation Detectors}.
\newblock {\em JINST}, 9:P06012, 2014.

\bibitem{Cheng:2015cen}
Yaping Cheng.
\newblock {Determination of Supernovae Direction with Reconstructed Positron
  Information}.
\newblock {\em PoS}, NEUTEL2015:067, 2015.

\end{thebibliography}
\end{document}